\documentclass[a4paper]{article} 
\usepackage[top=3cm,bottom=2cm,left=2cm,right=2cm]{geometry}
\usepackage[numbers,sort&compress]{natbib} 
\usepackage[utf8]{inputenc} 
\usepackage[T1]{fontenc}    
\usepackage{hyperref}       
\usepackage{url}            
\usepackage{booktabs}       
\usepackage{amsfonts}       
\usepackage{nicefrac}       
\usepackage{microtype}      
\usepackage{lipsum}
\usepackage{amsthm,amsmath,amssymb,bm}
\usepackage{mathrsfs}
\usepackage[pdftex]{graphicx}
\usepackage{rotating}
\usepackage{scalefnt}
\usepackage{longtable}
\usepackage{tcolorbox}
\usepackage{extarrows}
\usepackage{soul}
\usepackage{float}
\usepackage{braket,cleveref}

\newtheorem{theorem}{Theorem}

\newtheorem{lemma}{Lemma}

\usepackage{authblk}

\title{\bf Detecting new edge types in a temporal network model}

\author[1]{Wenjie Jia}
\author[2,3]{Manuel S. Mariani}
\author[2,4,$\dag$]{Linyuan L\"{u}}
\author[1,$\dag$]{Tao Jiang}

\affil[1]{\small School of Electronic Information and Communications and Wuhan National Laboratory for Optoelectronics, Huazhong University of Science and Technology, Wuhan 430074, P. R. China}
\affil[2]{\small Yangtze Delta Region Institute (Huzhou) and Institute of Fundamental and Frontier Sciences, University of Electronic Science and Technology of China,  Huzhou 313001, P. R. China}
\affil[3]{\small URPP Social Networks, University of Zurich, Zurich CH-8050, Switzerland}
\affil[4]{\small Beijing Computational Science Research Center, Beijing 100193, P. R. China}
\affil[$\dag$]{\small To whom correspondence should be addressed: linyuan.lv@uestc.edu.cn, taojiang@hust.edu.cn}

\date{}

\begin{document}
\maketitle

\begin{abstract}
Networks representing complex systems in nature and society usually involve multiple interaction types. These types suggest essential information on the interactions between components, but not all of the existing types are usually discovered. Therefore, detecting the undiscovered edge types is crucial for deepening our understanding of the network structure. Although previous studies have discussed the edge label detection problem, we still lack effective methods for uncovering previously-undetected edge types. Here, we develop an effective technique to detect undiscovered new edge types in networks by leveraging a novel temporal network model. Both analytical and numerical results show that the prediction accuracy of our method is perfect when the model networks’ time parameter approaches infinity. Furthermore, we find that when time is finite, our method is still significantly more accurate than the baseline.
\end{abstract}

\section{Introduction}
Complex systems in nature and society, including biology, transportation systems, computer science and social science, usually involve multiple interaction types leading the networks representing these systems to exhibit heterogeneous structures~\cite{boccaletti2014the,ahn2010link,clauset2008hierarchical,leskovec2010signed, Zlatic2011ComplexNW,battiston2020networks,Gemmetto2016MultiplexityAM,pilosof2017the}. 
Ordinarily, these different interaction types are represented by distinct edge labels. For example, in protein-protein interaction (PPI) networks, nodes represent proteins, edges connect pairs of interacting proteins, and the labels assigned on each edge indicate what types of interactions the edge represents. Without complete label information of edges in a network, it is impossible to fully understand the network's heterogeneous structure and properties, including robustness~\cite{Havlin2010CatastrophicCO,cencetti2019robustness} resilience~\cite{Gao2016UniversalRP} and dynamical properties~\cite{domenico2016the,arruda2018fundamentals,cencetti2017multiplex,cencetti2019diffusive} of these systems.

However, in many cases, we can only access a part of the complete edge label information, and there could exist {previously-undiscovered} interactions in the systems which are still undiscovered.  For instance, in PPI networks, the edge labels referring to the protein-protein interactions are often obtained from protein complex detection. Due to the limitation of detection techniques, such complex detection could only provide information on some specific interaction types, and edges of other interaction types {({\it i.e.} edges with previously-undiscovered labels) could exist. Because of this, there is an edge label detection problem: identify the edges that are the most likely to exhibit previously-undiscovered interaction types. Many research projects would benefit from the solutions to this problem. For example,} these techniques would help biologists speed up the discovering of new types of interactions between proteins and reduce the biological experiment costs.

The proposed edge label detection problem aims to detect edges with new labels from existing ones. The problem is different from previously-formulated problems that aim to predict old labels of sets of edges from the observed labeled edges, which has been widely studied in existing link annotation research~\cite{hric2016network,newman2016structure,ren2018structure}, including the sign prediction~\cite{leskovec2010predicting,song2015link,yuan2017negative,khodadadi2017sign,shahriari2016sign} and link prediction~\cite{lu2011link,lu2015toward,Parisi2018EntropybasedAT}.
We mathematically describe the proposed problem which has not been studied in existing research before, as follows.
Consider a network ${\bf G}$ whose structure ({\it i.e.} the set of nodes $\mathcal{V}$ and the set of edges $\mathcal{E}$) is fully observed, but the edge labels are only partially observed. Let $\mathcal{C}$ be the set of labels having been observed, and $\mathcal{E}_{label}\subseteq\mathcal{E}$ be the set of edges with labels in $\mathcal{C}$. The task of the new edge label detection is to find out a small edge set $\mathcal{E}'\subseteq\mathcal{E}$ in which every edge is likely to carry a new label $C'\not\in\mathcal{C}$ based on the known label information and the network structure. Unlike in existing research, we face new challenges in our problem: for a previously-undiscovered label, we know neither what it stands for nor the interacting features between it and other already-observed labels, making it seem hopeless to solve this problem only from the incomplete label information. Consequently, all the existing edge label prediction methods are not fit for the proposed problem. In other words, we can solve this problem only by random guessing currently.

Here, to overcome this barrier, we first propose a degradation-evolution network model. In this temporal network model, a network's structure is time-varying and allowed to mutate spontaneously by rewiring edges, and a potential energy model quantifies the degree of its susceptibility to mutation. The higher a network's potential energy is, the higher chance for it to have a different structure in the near future. We say a network evolves if its potential energy decreases and degrades if the potential energy increases. Then we consider this problem in the synthetic networks generated by this model and find that when the investigated networks enter into a stationary state, the networks admit a particular topological property, which enables us to make perfect detection for edges with new labels.  Next, we apply the newly developed detection method to a number of synthetic networks that are not in the stationary state and find that the method's accuracy is markedly higher than the accuracy of random guessing.

\begin{figure}
\includegraphics[width=10cm,angle=0]{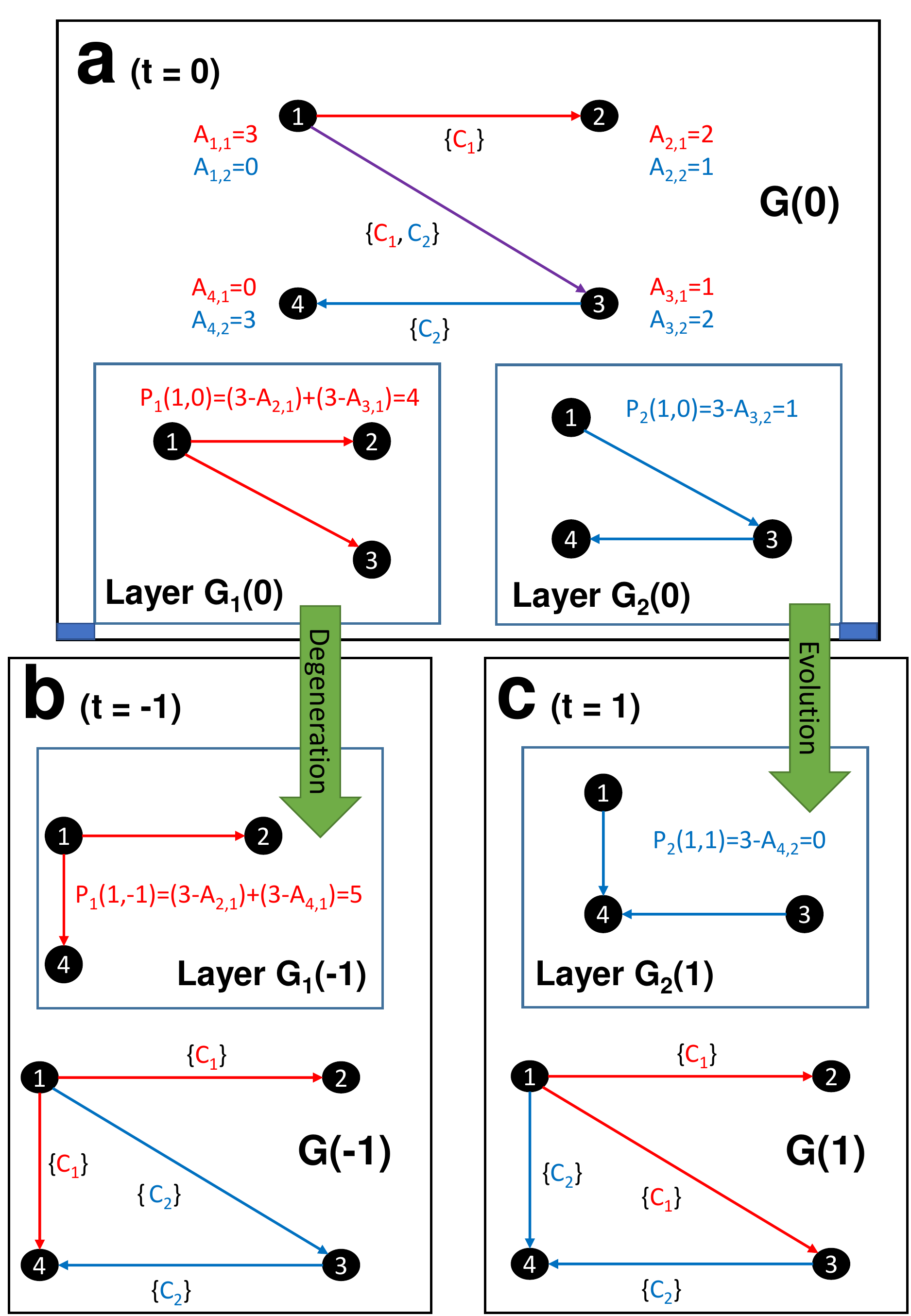}
\caption{An illustration of the  degradation-evolution network model. In {\bf a}, we initialize network ${\bf G}$ at time $t=0$ as a network consisting of $4$ nodes and $3$ directed edges. Each node is assigned a pair of $1$-attractiveness and $2$-attractiveness. The three edges are tagged by two kinds of labels ({\it i.e.} $C_1$ and $C_2$).
In {\bf b}, the node with index $1$ increases its potential energy with respect to the first layer by rewiring its out-edges in layer ${\bf G}_1(0)$, making ${\bf G}$ degrade into ${\bf G}(-1)$. In {\bf c}, the node with index $1$ rewires its out-edges in layer ${\bf G}_2(0)$ making $P_2(1,1)<P_2(1,0)$ and ${\bf G}$ evolve into ${\bf G}(1)$.}
\label{fig.1}
\end{figure}

\section{The model}
The synthetic networks studied in this letter are generated by a temporal degradation-evolution network model, which is introduced as follows. 
Let $t$ be the time. We use $t<0$, $t=0$ and $t>0$ to denote the past, the present and the future, respectively. At the present time ({\it i.e.} $t=0$), we initialize ${\bf G}={\bf G}(0)$ to be an arbitrary network with $n$ nodes and $m$ edges with labels in $\mathcal{C}$, where $\mathcal{C}=\{C_1,C_2,\cdots,C_h\}$ represents all the labels that can be observed in the whole course of ${\bf G}$'s changing process.
Let ${\bf G}(t)=(\mathcal{V}(t),\mathcal{E}(t))$ be a temporal network with $\mathcal{V}(t)=\{v_1,v_2,\cdots,v_n\}$, for time $t=-\infty,\cdots,-1,0,1,\cdots,+\infty$. For an edge $v_i\rightarrow v_j$ in ${\bf G}(t)$, we employ notation $\mathcal{C}(i,j,t)$ to denote the label set associated with it. 
We assume that every edge in networks should be assigned at least one label. Therefore, there exists an edge from $v_i$ to $v_j$ at time $t$ if and only if $\mathcal{C}(i,j,t)\neq\emptyset$.

Regarding edges with the same label in ${\bf G}$ as the components of a layer of network ${\bf G}$, we can divide ${\bf G}$ into $h$ different layers. 
Specifically, we define the $l$-th layer of ${\bf G}(t)$, denoted by ${\bf G}_{l}(t)=(\mathcal{V}_{l}(t),\mathcal{E}_{l}(t))$, to be the subnetwork consisting of all the edges in $\mathcal{E}(t)$ with label $C_l$ and all the nodes involved in these edges (see Fig. \ref{fig.1}{\bf a}).
Inspired by the attractiveness model~\cite{dorogovtsev2000structure}, we assume: (1) for $l\in\{1,2,\cdots,h\}$, every node $v_{i}$ in ${\bf G}(t)$ is assigned with an attractiveness $A_{i,l}\geq 0$, called $v_{i}$'s $l$-attractiveness or attractiveness associated with layer ${\bf G}_{l}(t)$ (see Fig. \ref{fig.1}{\bf a}); (2) in each layer of the system, a node always intends to connect to nodes with high attractiveness associated with this layer, and it can rewire its out-edges in the layer to better fulfill this intention.

For node $v_i$ we define its potential energy with respect to the $l$-th layer at time $t$ to be
\begin{equation}\label{eq_potential_energy}
P_l(i,t)=\sum_{j=1}^{n}(A_{max,l}-A_{j,l})\chi_{\mathcal{E}_{l}(t)}(v_i\rightarrow v_j),
\end{equation}
where $A_{max,l}=\max_{v_i\in\mathcal{V}}A_{i,l}$ and $\chi_{\mathcal{E}_{l}(t)}(v_i\rightarrow v_j)=1$ if $v_i\rightarrow v_j\in\mathcal{E}_{l}(t)$; otherwise $\chi_{\mathcal{E}_{l}(t)}(v_i\rightarrow v_j)=0$.
We employ $P_l(i,t)$ to describe how eager node $v_i$ is to rewire its out-edges in ${\bf G}_{l}(t)$ to connect to nodes with higher $l$-attractiveness at time $t$ (see Fig. \ref{fig.1}{\bf a}).
Further, we define $v_i$'s potential energy and the system's potential energy at time $t$ to be
$P(i,t)=\sum_{l=1}^{h}P_l(i,t)$ and $P({\bf G},t)=\sum_{i=1}P(i,t)$, respectively. A node's higher potential energy means the stronger desire for this node to rewire its out-edges, and the higher potential energy of a system indicates a more structurally unstable state of this system.

We introduce an evolution mechanism: at each time $t>0$, a node rewires some of its out-edges in a layer of ${\bf G}(t)$ and then time $t$ increases by $1$, such that $P({\bf G},t+1)\leq P({\bf G},t)$ (see Fig. \ref{fig.1}{\bf c}). In addition, we assume that there also exists a degradation mechanism. That is, at each time $t<0$, a node rewires some of its out-edges in a layer of ${\bf G}(t)$ and then time $t$ decreases by $1$, such that $P({\bf G},t-1)\geq P({\bf G},t)$ (see Fig. \ref{fig.1}{\bf b}).

\section{Topological property}
In the rest of this letter, we always let ${\bf G}(t)$ be a network observed at time $t$ with $\mathcal{V}(t)$ and $\mathcal{E}(t)$ denoting its nodes and edges. Let $E\subseteq\mathcal{E}(t)$ be a non-empty edge set.
We use notation $\lceil\,E\,\rceil_{src}$ ($\lceil\,E\,\rceil_{tar}$) to denote the set consisting of all the source (target) nodes of edges in $E$.
We say $v_i$ is a $(2,{\bf G}(t))$-follower of $v_j$ in ${\bf G}(t)$ if the shortest simple path (a simple path is a path without repeated nodes) from $v_i$ to $v_j$ in ${\bf G}(t)$ is $2$.
Assume that different nodes have different $l$-attractiveness for any $l\in\{1,2,\cdots,h\}$. Then we obtain the following lemma (see its derivation in Appendix A).
\begin{lemma}\label{lemma1}
Let $E$ be an arbitrary nonempty subset of $\mathcal{E}(t)$.
At time $|t|=+\infty$, if all the edges in $E$ lie in the same layer, then there always exists a node in $\lceil\,E\,\rceil_{tar}$ having no $(2,{\bf G}(t))$-follower in $\lceil\,E\,\rceil_{src}$.
\end{lemma}

We say $E\subseteq\mathcal{E}(t)$ has the Delta-property, denoted by $\Delta(E)=1$, if 
$\delta(E)=\emptyset$, where $\delta(E)$ can be obtained by implementing the following procedures:
(1) select a node $v$ from $\lceil\,E\,\rceil_{tar}$ which has no $(2,{\bf G}(t))$-follower in $\lceil\,E\,\rceil_{src}$; 
(2) remove all the edges whose target node is $v$ from $E$;
(3) repeat (1)-(2) until no more removal is possible;
and (4) set the remaining edge set to be $\delta(E)$. 
In addition, we define $\Delta(E)=0$, if $\delta(E)\neq\emptyset$.
Then we have the following results (see derivations in Appendices B and C).

\begin{lemma}\label{lemma2}
Implementing the removal operations on $E$ introduced above, we obtain $\{E_1,E_2,\cdots,E_s\}$ and $\delta(E)$, where $E_i$ denotes the edge set removed from $E$ in the $i$-th removal operation, for $i=1,2,\cdots,s$.
Let $\delta(E)=E_{s+1}$.
Then we have:
(1) $E_i\not=\emptyset$, for $i=1,2,\cdots,s$;
(2) $E_i\cap E_j=\emptyset$, for $1\leq i<j\leq s+1$;
(3) $\cup_{i=1}^{s+1}E_i=E$;
(4) if $\delta(E)\neq\emptyset$, then $|\delta(E)|\geq 2$.
\end{lemma}

\begin{lemma}\label{lemma3}
Let $E$ be a non-empty subset of $\mathcal{E}(t)$.
Let $\delta^{(1)}(E)$ and $\delta^{(2)}(E)$ be two subsets of $E$ obtained by implementing the removal procedures introduced above.
Then we have $\delta^{(1)}(E)=\delta^{(2)}(E)$.
\end{lemma}

Lemma \ref{lemma3} shows that mapping $\Delta$ is well-defined. Further, we obtain our main theoretical result (see its derivation in Appendix D). 

\begin{theorem}\label{thm}
When $t=\pm\infty$, for an edge set $E\subseteq\mathcal{E}(t)$ if $\Delta(E)=0$, then one must have all the edges composing $E$ can not share a common label. 
\end{theorem}

The above theorem shows that when a network is fully evolved or fully degraded, its multilayer structure must follow a special topological property. Specifically, we can apply this result to judge whether these edges can share common labels for any given set of edges. In the following, we show how to utilize this result to detect previously-undiscovered edge labels.

\section{Detection method}
Based on Theorem \ref{thm}, we derive a method to tackle the new label detection problem. Let $t=\pm\infty$, assume that in ${\bf G}(t)$, only partial edges have known label information. Specifically, let $\mathcal{C}(t)=\{C_{t_1},C_{t_2},\cdots,C_{t_s}\}$ denote all the edge label observed at time $t$, $s=s(t)$ be an integer, and $E_{t_l}\in\mathcal{E}_{t_l}(t)$ consist of all the edges which are observed with label $C_{t_l}$ at time $t$, for $1\leq l\leq s(t)$. 
If $C\in\mathcal{C}$ is an edge label with $C\not\in\mathcal{C}(t)$, then we call $C$ is a previously-undiscovered edge label at time $t$.
Let $e$ denote an edge in $\mathcal{E}(t)$ satisfying that $\Delta(\{e, e_{t_l}\}) =0$ for any $e_{t_l}\in E_{t_l}$ and $l\in\{1,2,\cdots,s(t)\}$. Then, according to Theorem \ref{thm}, we have $e$ must own a new edge label. 
Assembling all of such edges, we obtain an edge set $E'$ in which every edge has at least one new label.
Our theory shows that this detection method's accuracy for $|t|=+\infty$ is perfect ($100\%$). 
For $|t|<+\infty$, the accuracy of this detection method is case-dependent.

In the rest of this letter, we focus on one of the most straightforward cases of our main problem. Let ${\bf G}=(\mathcal{V},\mathcal{E})$ be a network, $C_1$ be an already-observed edge label, $\mathcal{E}_1$ be the set consisting of all the edges in ${\bf G}$ carrying label $C_1$, and $E_1\subseteq\mathcal{E}_1$ be an edge set consisting of $n_1$ edges with label $C_1$. Our goal is to find out a small number of edges with previously-undiscovered labels based on $E_1$ and ${\bf G}$'s topology. To solve this problem, we assign each edge $e$ in $\mathcal{E}$ a score $\nabla(e|E_1)$ with
$
\nabla(e|E_1)=|\{e'\in E_1|\Delta(\{e,e'\})=0\}|.
$
Then take the edges with the largest nonzero scores as the algorithm's output. We use notation $E_1$-D-Top and $E_1^{*}$ to denote the corresponding algorithm and its output, respectively. Note that for any edge $e$, $\nabla(e|E_1)$ is an integer and $\nabla(e|E_1)\leq |E_1|=n_1$. Thus, for small $n_1$, such as $n_1=1,2,3$, there would be a small difference in the edges' scores, which could impair the performance of our method. For small $n_1$, we further require that the detected/output edges by $E_1$-D-Top should have a score of $n_1$ ({\it i.e.} $\nabla(e|E_1)=n_1$ for any $e\in E_1^{*}$).

Two standard metrics are used to quantify the accuracy of detection algorithms: Precision~\cite{herlocker2004evaluating} and {\it area under the receiver operating characteristic curve} (AUC)~\cite{hanley1982the}. Assume that in a detected edge set $E^{*}_1$ consisting of $n_2$ edges, there are $n'$ edges are right ({\it i.e.} there are $n'$ edges are with previously-undiscovered labels), then the Precision of this algorithm is $n'/n_2$.  Here, we use $\omega({\bf G}|E_1)$ to denote the Precision of algorithm $E_1$-D-Top.  Higher Precision means higher detection accuracy. Note that for a given edge set $E_1\subseteq\mathcal{E}_1$ with $|E_1|=n_1$, the performance of algorithm $E_1$-D-Top is closely related to the probability that an arbitrary edge with label $C_2$ gets a larger score than another arbitrary edge with label $C_1$, which can be quantified by AUC~\cite{hanley1982the}. To measure the AUC, denoted by $\lambda({\bf G}|E_1)$, we can make $N$ independent comparisons: at each time, we randomly pick an edge with previously-undiscovered labels and an edge without previously-undiscovered labels to compare their scores. If there are $N'$ times the edge with undiscovered labels obtaining a higher score and $N''$ times they have the same score, then the AUC value is 
$\lambda({\bf G}|E_1)=(N'+0.5N'')/N$~\cite{lu2011link}. 
If all the scores are generated from an independent and identical distribution, the AUC value should be about $0.5$. Therefore, the degree to which the value exceeds $0.5$ indicates how much better the algorithm performs than random guessing.
In this letter, we only consider networks with small numbers of nodes and small numbers of edges. In this scenario, we can run through all possible combinations of edges without previously-undiscovered labels and edges with previously-undiscovered labels to measure the AUC of the network.

We are interested in our method's accuracy in detection new edge labels in ${\bf G}$, when we are given an edge set consisting of $n_1$ edges arbitrarily picked from $\mathcal{E}_1$. We use notation $n_1$-D-Top to represent algorithm $E_1$-D-Top,  where $E_1$ is an arbitrary subset of $\mathcal{E}_1$ with $n_1$ elements. We denote $\omega_{n_1}({\bf G})$ and $\lambda_{n_1}({\bf G})$ as the Precision and the AUC of $n_1$-D-Top, respectively. Then we obtain 
\begin{equation}\label{eq_result1}
\omega_{n_1}({\bf G})=\frac{1}{\binom{|\mathcal{E}_1|}{n_1}}
\sum_{E_1\subseteq\mathcal{E}_1,|E_1|=n_1}\omega({\bf G}|E_1)
\end{equation}
and
\begin{equation}\label{eq_result2}
\lambda_{n_1}({\bf G})=\frac{1}{\binom{|\mathcal{E}_1|}{n_1}}
\sum_{E_1\subseteq\mathcal{E}_1,|E_1|=n_1}\lambda({\bf G}|E_1),
\end{equation}
where $\binom{|\mathcal{E}_1|}{n_1}=|\mathcal{E}_1|!/[n_1!(|\mathcal{E}_1|-n_1)!]$ is a combination.

For a given network ${\bf G}$ whose structure varies over time,  we are concerned about the accuracy of the proposed algorithms applied to ${\bf G}$ at present and curious about both what happened to their performance in the past and what their performances will become in the future. Let ${\bf G}$ be some network generated by our proposed degradation-evolution model. Denoting the present time as $t=0$, we can rewrite ${\bf G}$ as ${\bf G}(0)$. By our degradation-evolution model, we obtain a family of networks $\{{\bf G}(t)\}_{-\infty}^{+\infty}$, which depicts the whole course of ${\bf G}$'s evolution. To study the overall performance of the detection methods, we introduce another parameter $\nu$, which is given by $\nu=\nu(t)=[P({\bf G},t)-P_{min}({\bf G})]/[P_{max}({\bf G})-P_{min}({\bf G})]$, where $P_{max}({\bf G})$ and $P_{min}({\bf G})$ denote the supremum and infimum of $P({\bf G},t)$, respectively. Parameter $\nu$ ranges from $0$ to $1$ and describes the evolution degree of ${\bf G}$: the nearer $\nu$ approaches to $0$, the more stable the structure of the network is. Then we can rewrite $\{{\bf G}(t)\}_{-\infty}^{+\infty}$ as $\{{\bf G}[\nu]\}_{\nu\in[0,1]}$ and ${\bf G}={\bf G}[\nu_0]$, where $\nu_0=\nu(0)$. Then the average Precision over time ($\bar{\omega}_{n_1}({\bf G})$) and the average AUC over time ($\bar{\lambda}_{n_1}({\bf G})$) of $n_1$-D-Top applied to ${\bf G}$ can be calculated by
\begin{equation}\label{eq_result3}
\bar{\omega}_{n_1}({\bf G})=\int_{0}^{1}\omega_{n_1}({\bf G}[\nu])f(\nu)d\nu
\end{equation}
and
\begin{equation}\label{eq_result4}
\bar{\lambda}_{n_1}({\bf G})=\int_{0}^{1}\lambda_{n_1}({\bf G}[\nu])f(\nu)d\nu
\end{equation}
respectively, where $f(\nu)$ refers to the probability density function of $\nu$. 
In this letter, we assume that $\nu$ subjects to a uniform distribution.

We study the performance of $n_1$-D-Top applied to randomly generated networks by the degradation-evolution model. For a given  $\bm{\theta}$ and a given $\nu\in[0,1]$, the Precision and AUC of $n_1$-D-Top applied to a random network ${\bf G}$ which is generated by the proposed model under a specific configuration given by $\bm{\theta}$ and admits ${\bf G}={\bf G}[\nu]$,  are represented by $\omega_{n_1,\bm{\theta}}(\nu)$ and $\lambda_{n_1,\bm{\theta}}(\nu)$, and can be calculated by
\begin{equation}\label{eq_result5}
\omega_{n_1,\bm{\theta}}(\nu)=\frac{1}{M}\sum_{i=1}^{M}\omega_{n_1}({\bf G}_i)
\end{equation}
and
\begin{equation}\label{eq_result6}
\lambda_{n_1,\bm{\theta}}(\nu)=\frac{1}{M}\sum_{i=1}^{M}\lambda_{n_1}({\bf G}_i)
\end{equation}
where ${\bf G}_i$ is a random network generated by the degradation-evolution model with specific parameters $\bm{\theta}$ and admitting ${\bf G}_i={\bf G}_i[\nu]$, for $i=1,2,\cdots,M$.
Finally, the Precision and AUC of $n_1$-D-Top applied to a randomly generated network by the model with specific parameter settings $\bm{\theta}$, can be represented as 
\begin{equation}\label{eq_result7}
\omega_{n_1,\bm{\theta}}=\int_{0}^{1}\omega_{n_1,\bm{\theta}}(\nu)f(\nu)d\nu
\end{equation}
and
\begin{equation}\label{eq_result8}
\lambda_{n_1,\bm{\theta}}=\int_{0}^{1}\lambda_{n_1,\bm{\theta}}(\nu)f(\nu)d\nu.
\end{equation}
It follows from Eqs. (\ref{eq_result3})--(\ref{eq_result8}) that
\begin{equation}\label{eq_result9}
\omega_{n_1,\bm{\theta}}=\frac{1}{M}\sum_{i=1}^{M}\bar{\omega}_{n_1}({\bf G}_i)
\end{equation}
and
\begin{equation}\label{eq_result10}
\lambda_{n_1,\bm{\theta}}=\frac{1}{M}\sum_{i=1}^{M}\bar{\lambda}_{n_1}({\bf G}_i)
\end{equation}
where ${\bf G}_i$ is a network randomly generated by the model with specific parameter settings $\bm{\theta}$, for $i=1,2,\cdots,M$. By Eqs. (\ref{eq_result1})--(\ref{eq_result6}), (\ref{eq_result9}) and (\ref{eq_result10}), we can readily investigate the Precision and AUC of $n_1$-D-Top applied to synthetic networks in practice.

\section{Experimental results and discussion}
Given a network ${\bf G}$ with $\mathcal{V}=\{v_1,v_2,\cdots,v_n\}$ and $|\mathcal{E}|=m$, we aim to detect the previously-undiscovered labels in ${\bf G}$. We assume that ${\bf G}$ is generated by the degradation-evolution model under following configurations: (1) there are totally two edge labels, $C_1$ and $C_2$, which can be observed in ${\bf G}$; (2) label $C_1$ is the already-observed label and $C_2$ is the previously-undiscovered one; (3) every edge has a unique edge label, and the the percentage of edges with undiscovered label $C_2$ is $\alpha$; and (4) node $v_i$'s $1$-attractiveness $A_{i,1}$ is $n-i$ and $2$-attractiveness $A_{i,2}$ is $i-1$, for $i=1,2,\cdots,n$. Let $\bm{\theta}$ represent these configurations.
By the degradation-evolution network model, we obtain a family of networks $\{{\bf G}(t)\}_{-\infty}^{+\infty}$. For any $t$, every edge in ${\bf G}(t)$ has a unique edge label. According to the rewiring procedures introduced before, we have ${\bf G}(t)$ always consists of $n$ nodes and $m$ edges.  
In the following, we investigate the Precision and AUC of $n_1$-D-Top applied to ${\bf G}(t)$, for $-\infty<t<+\infty$. Note that the random guessing has a Precision of $\alpha$ and an AUC of $0.5$ in this case.

\begin{figure}
\includegraphics[width=15cm,angle=0]{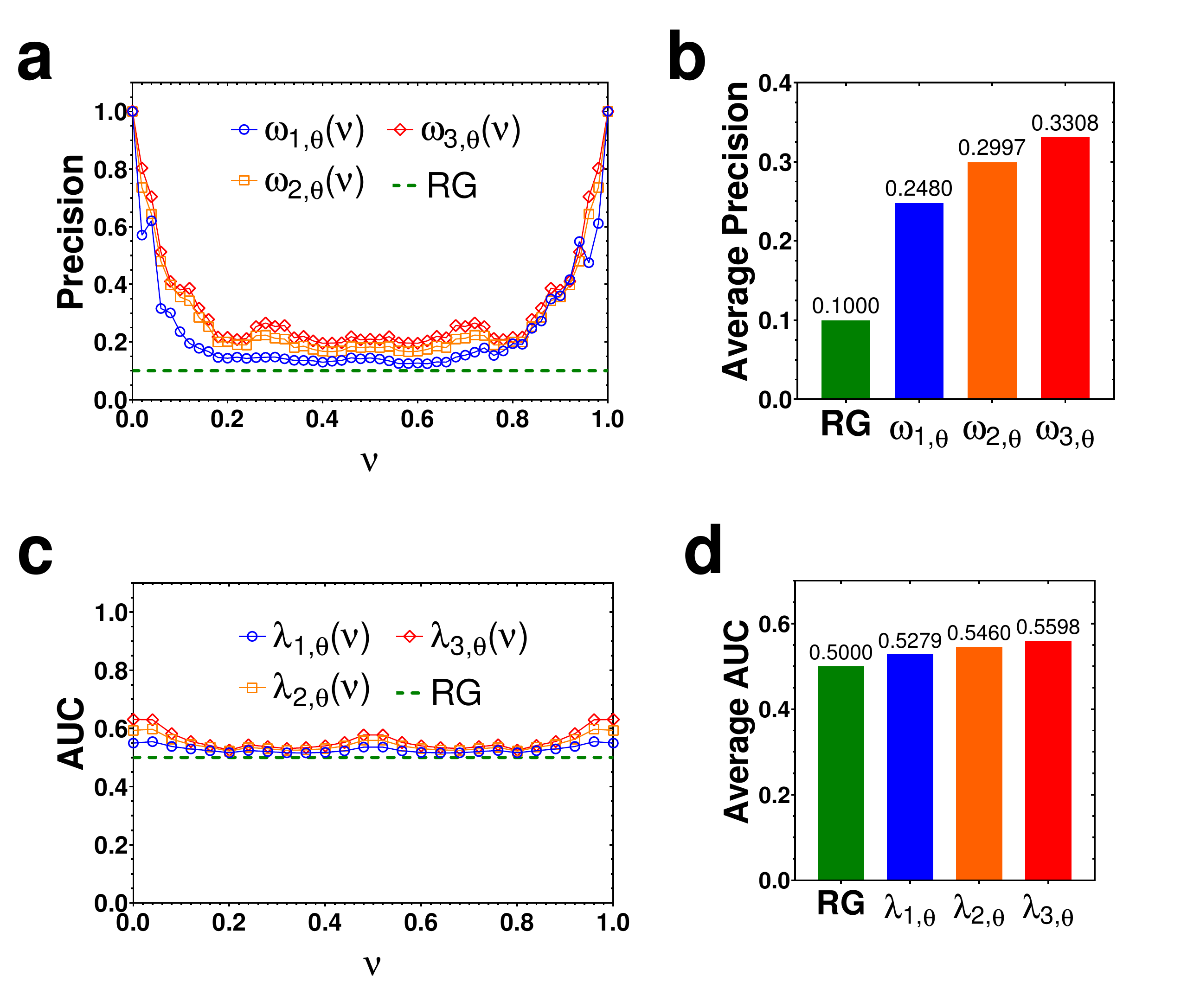}
\caption{The accuracy of $n_1$-D-Top as functions of $\nu$ in detecting rare previously-unobserved labels (small $n_1$).
The parameter settings are as follows: $n=20$, $m=100$, $\alpha=0.1$ and $n_1\in\{1,2,3\}$.
RG represents the random guessing. Each result is averaged by over $100$ independent implementations.}
\label{fig.2}
\end{figure}

\begin{figure}
\includegraphics[width=15cm,angle=0]{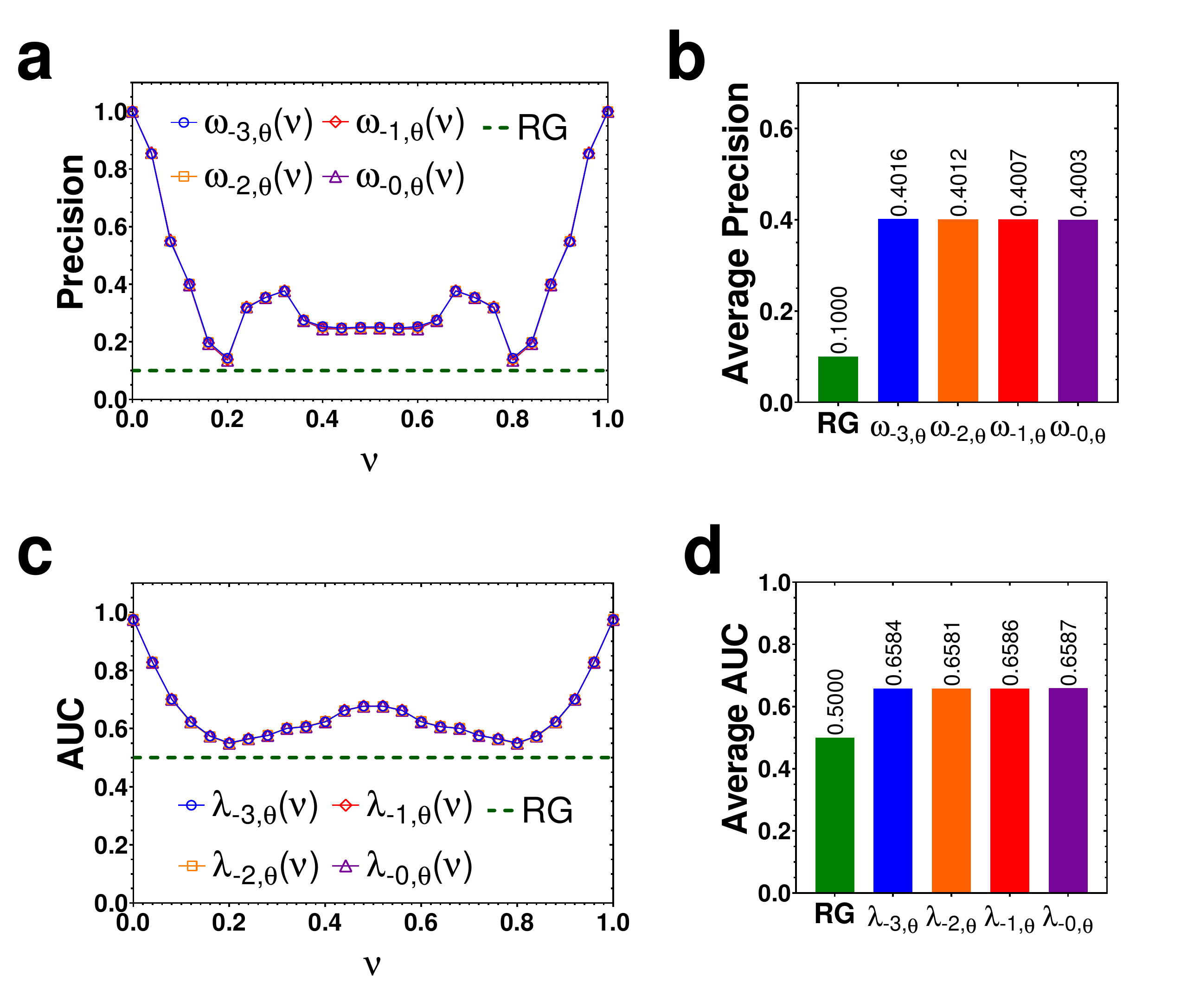}
\caption{The accuracy of $n_1$-D-Top as functions of $\nu$ in detecting rare previously-unobserved labels (large $n_1$).
The parameter settings are as follows: $n=20$, $m=100$, $\alpha=0.1$ and $n_1\in\{-3,-2,-1,-0\}$.
RG represents the random guessing. Each result is averaged by over $100$ independent implementations.}
\label{fig.3}
\end{figure}

We study the performance of $n_1$-D-Top with small $n_1$ in detecting rare unobserved labels.
Figure 1 plots the Precision and the AUC of $1$-D-Top, $2$-D-Top and $3$-D-Top as functions of $\nu$ when $\alpha=0.1$. From Figure \ref{fig.2}a, we find that when $\nu\in\{0,1\}$ ({\it i.e.} $|t|=+\infty$), the algorithms always gain the perfect Precision ($100\%$), which is consistent with our theoretical results. When $\nu$ is in the middle of the interval $[0,1]$ (for instance, $\nu\in[0.2,0.8]$), the Precision of each algorithm is stable, while $\nu$ approaches to $0$ or $1$, the Precision will increase sharply. From Figure \ref{fig.2}b, we find that the average Precision of $1$-D-Top, $2$-D-Top and $3$-D-Top ({\it i.e.} $\omega_{1,\bm{\theta}}$, $\omega_{2,\bm{\theta}}$ and $\omega_{3,\bm{\theta}}$) are $0.2480$, $0.2997$ and $0.3308$, respectively. We conclude that $1$-D-Top, $2$-D-Top and $3$-D-Top are valid and effective, since they all outperform random guessing, and improve the Precision of random guessing ($10\%$) by $148.0\%$, $199.7\%$ and $230.8\%$, respectively. Moreover, as shown in Figure \ref{fig.2}b that the Precision of $n_1$-D-Top increases as $n_1$ increases, showing that detection based on more edges with observed labels would get better performance. From Figure \ref{fig.2}c and Figure \ref{fig.2}d, we see that the average AUC of $1$-D-Top, $2$-D-Top and $3$-D-Top ({\it i.e.} $\lambda_{1,\bm{\theta}}$, $\lambda_{2,\bm{\theta}}$ and $\lambda_{3,\bm{\theta}}$) are close to $0.5$ showing that $n_1$-D-Top has a poor performance in AUC when $n_1$ is small, which is consistent with our previous judgment (see ``Detection method" section).

We consider the performance of $n_1$-D-Top with large $n_1$ in detecting rare unobserved labels.
We use notation $-n_1$-D-Top to represent $(m_1-n_1)$-D-Top, where $m_1$ denote the total number of edges with label $C_1$. For example, $-0$-D-Top refers to the proposed detection method based on all the edges with $C_1$. Figure \ref{fig.3} plots the Precision and the AUC of $-n_1$-D-Top as functions of $\nu$ in the case of $\alpha=0.1$ for $n_1\in\{0,1,2,3\}$. From Figures \ref{fig.3}a and \ref{fig.3}c, we find the four detection algorithms have almost the same accuracy, indicating that they have almost achieved the upper bound of the proposed method's performance. Figures \ref{fig.3}b and \ref{fig.3}d demonstrate that the proposed method can improve the Precision and AUC of random guessing by $\geq300\%$ and  $\geq31\%$, respectively.

\begin{figure}
	\includegraphics[width=15cm,angle=0]{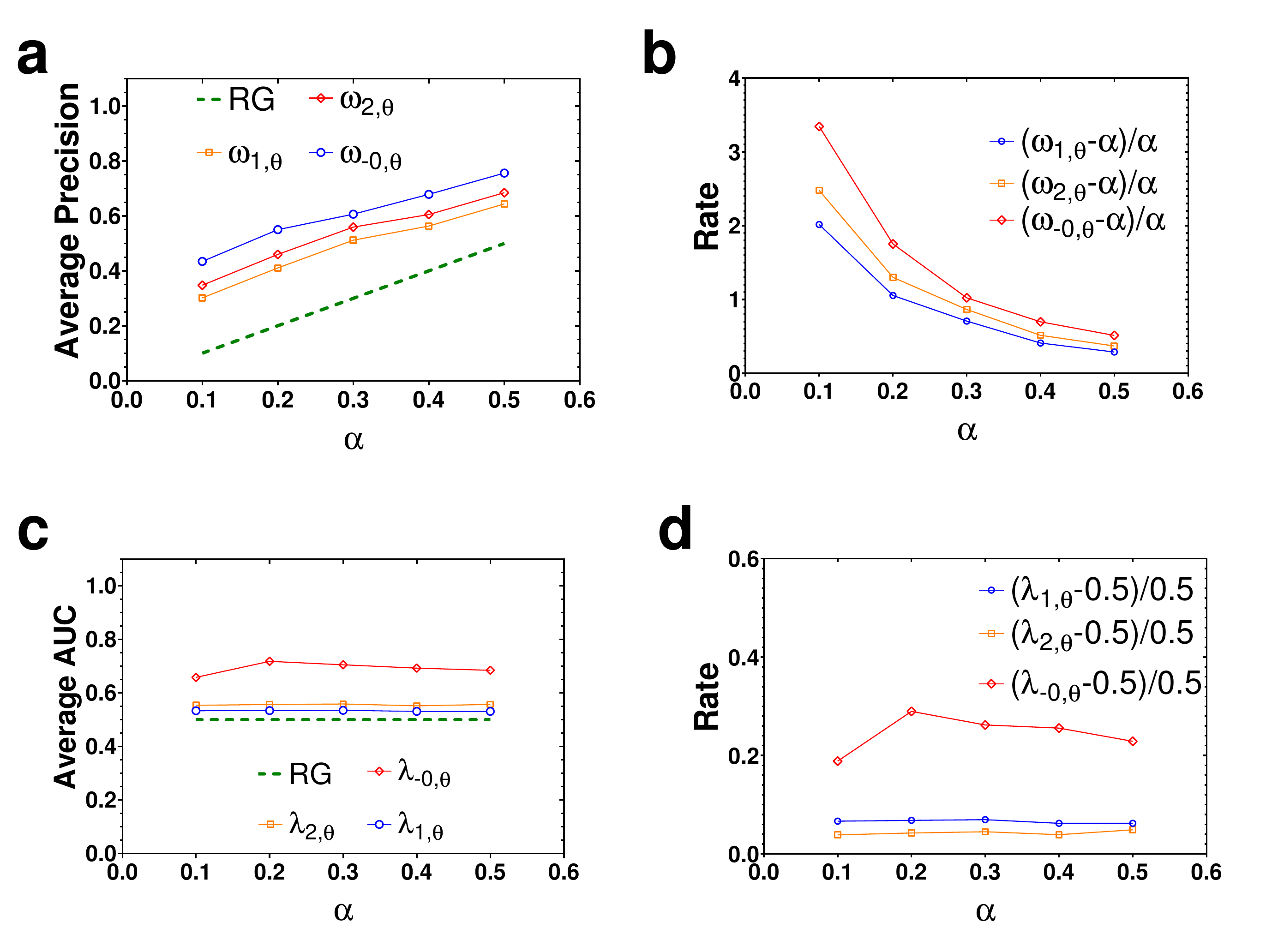}
	\caption{The average accuracy of $n_1$-D-Top as functions of $\alpha$ in detecting previously-unobserved labels.
	The parameter settings are as follows: $n=20$, $m=100$, $\alpha\in\{0.1,0.2,\cdots,0.5\}$ and $n_1\in\{-0,1,2\}$.
	RG represents the random guessing. Each result is averaged by over $100$ independent implementations.}
	\label{fig.4}
	\end{figure}

When the unobserved label is not rare, we show the performance of $n_1$-D-Top in Figure \ref{fig.4}.  
As shown above, $1$-D-Top and $-0$-D-Top are the algorithms with the worst performance and best performance, and their performance outlines the feasible region of the accuracy of the proposed method. It can be seen from Figures \ref{fig.4}a and \ref{fig.4}c that as the rarity of the unobserved label ($\alpha$) increases, the Precision of our method also increases linearly, while the AUC usually holds steady. Interestingly, from Figure \ref{fig.4}b, we find that the rarer the unobserved label is, the more largely our method improves the Precision of random guessing.

\section{Conclusions} 
In this letter, we propose a new edge label detection problem in which we {aim to} find out a small set of edges with {previously-unobserved} labels. On the one hand, as the target labels are previously-unobserved, we can not utilize the interacting features between previously-unobserved labels and the already-observed ones, making this problem challenging. On the other hand, it is essential to solve this problem since its solutions would benefit researchers in mining new features of a wide range of datasets, for instance, to discover new interactions between proteins. We propose a temporal directed network model and develop an effective detection method for synthetic networks generated by the proposed model, which entirely takes advantage of networks' topological properties. We focus on one of the most straightforward cases of the target problem: detecting the unobserved label in networks in which one label is observed, and one is unobserved. Applying our method to tackle this particular problem in synthetic networks generated by our proposed model, we find that our detection method is effective and has much better performance than the baseline. More complex cases of the original problem, for example, detecting the previously-unobserved labels when at least two labels have been observed, are still awaiting further exploration. Our method can be applied to real-world networks as well, which suggests some further directions: investigating the performance of the proposed method in real networks, exploring more methods that yield better performance for detecting new edge labels in both synthetic networks and real-world networks, and so on.

\section*{Acknowledgments}
This work is supported by the National Natural Science Foundation of China (Grant Nos. 61673150, 11622538).
LL acknowledges the Science Strength Promotion Programme of UESTC, Chengdu.

\section*{Appendix A: Proof of Lemma 1}
In the limit $t\rightarrow +\infty$ ($-\infty$), we have 
\begin{equation}\label{eq_lemma1}
P({\bf G},t)=P_{max}({\bf G})\ \ (P_{min}({\bf G})),
\end{equation}
where ${\bf G}={\bf G}(t)$, $P_{max}({\bf G})$ and $P_{min}({\bf G})$ denote the supremum and infimum of $P({\bf G},t)$, respectively. Assume that all the edges in $E$ lie in ${\bf G}_{l}(t)$. Note that all the nodes in $\lceil\,E\,\rceil_{tar}$ have different $l$-attractiveness. Without loss of generality, let $v_{1}$ ($v_{2}$) denote the node with the largest (smallest) $l$-attractiveness in $\lceil\,E\,\rceil_{tar}$. In the following, we show that $v_{1}$ ($v_{2}$) has no $(2,{\bf G}(t))$-follower in $\lceil\,E\,\rceil_{src}$ when $t=+\infty$ ($t=-\infty$) through the reverse proving.
Assume $v_j\in \lceil\,E\,\rceil_{src}$ is a $(2,{\bf G}(t))$-follower of $v_{1}$ ($v_{2}$).
Obviously, we have $v_j\not\rightarrow v_{1}$ ($v_j\not\rightarrow v_{2}$). By $v_j\in\lceil\,E\,\rceil_{src}$, there is $v_k\in\lceil\,E\,\rceil_{tar}$ such that $k\neq 1$ ($k\neq 2$) and $v_j\rightarrow v_k\in E$. Then we have $C_l\in\mathcal{C}(j,k,t)$ and $A_{k,l}<A_{1,l}$ ($A_{k,l}>A_{2,l}$).
In ${\bf G}_{l}(t)$, change $v_j\rightarrow v_k$ to $v_j\rightarrow v_1$ and let $t$ increase (decrease) by $1$.
By Eq. (\ref{eq_potential_energy}), we have $P_l(j,t+1)-P_j(j,t)=A_{1,l}-A_{k,l}>0$ ($P_l(j,t-1)-P_j(j,t)=A_{2,l}-A_{k,l}<0$) Then one has $P({\bf G},t)<P_{max}({\bf G})$ ($P({\bf G},t)>P_{min}({\bf G})$) which contradicts Eq. (\ref{eq_lemma1}).
Finally, we conclude that there always exists a node in $\lceil\,E\,\rceil_{tar}$ having no $(2,{\bf G}(t))$-follower in $\lceil\,E\,\rceil_{src}$ when $t=+\infty$ ($t=-\infty$).

\section*{Appendix B: Proof of Lemma 2}
We have $E_1,E_2,\cdots,E_{n}$ are pairwise disjoint non-empty subsets of $E$, and $\delta(E)$ is the remaining set.
Then, we have $\cup_{i=1}^{n+1}E_i=E$.
We show that $|\delta(E)|\neq 1$ by the reverse proving. 
Assume $|\delta(E)|=1$. Without loss of generality, we assume $\delta(E)=\{v_1\rightarrow v_2\}$.
Then $\lceil\delta(E)\rceil_{src}=\{v_1\}$ and $\lceil\delta(E)\rceil_{tar}=\{v_2\}$.
Note that $v_2$ has no $(2,{\bf G})$-follower in $\lceil\delta(E)\rceil_{src}$.
According to the removal operations, $v_1\rightarrow v_2$ can be removed. 
Thus, we have $|\delta(E)|\neq 1$.
In the following, we construct a network ${\bf G}=(\mathcal{V},\mathcal{E})$ and a set $E\subseteq\mathcal{E}$ with $|\delta(E)|=2$.
Let $\mathcal{V}=\{v_1,v_2,v_3,v_4\}$ and $\mathcal{E}=\{v_1\rightarrow v_2,v_1\rightarrow v_3,v_3\rightarrow v_1,v_3\rightarrow v_4\}$.
Let $E=\{v_1\rightarrow v_2,v_3\rightarrow v_4\}$.
We have $\lceil E\rceil_{src}=\{v_1,v_3\}$ and $\lceil E\rceil_{tar}=\{v_2,v_4\}$.
Note that $v_1$ is a $(2,{\bf G})$-follower of $v_4$ and $v_3$ is a $(2,{\bf G})$-follower of $v_2$.
According to the removal operations, we have $\delta(E)=E$ and $|\delta(E)|=|E|=2$.
To sum up, we have if $\delta(E)\neq\emptyset$, then $|\delta(E)|\geq 2$.

\section*{Appendix C: Proof of Lemma 3}
Let ${\bf G}={\bf G}(t)$, $k\in\{1,2,\cdots,s\}$ and $E^{(1)}_k$ denote the set consisting of all the edges removed from $E$ in the $k$-th removal operation, and $\delta^{(1)}(E)$ be the remaining set.
According to Lemma 2 (3), we have 
\begin{equation}\label{eq_lemma_delta_new_b_1}
\delta^{(1)}(E)\cup(\cup_{l=k}^{s}E^{(1)}_k)=E.   
\end{equation}

{Case 1}: $|\delta^{(2)}(E)|>0$. 
First, we show $\delta^{(2)}(E)\subseteq\delta^{(1)}(E)$ through the reverse proving.
Assume $\delta^{(2)}(E)\not\subseteq\delta^{(1)}(E)$.
It follows Eq.  (\ref{eq_lemma_delta_new_b_1}) that 
$\delta^{(2)}(E)\cap(\cup_{k=1}^{s}E^{(1)}_k)\neq \emptyset$.
Let $l$ be the smallest integer, such that $\delta^{(2)}(E)\cap E^{(1)}_l\neq\emptyset$.
Then we have 
$$
\delta^{(2)}(E)\subseteq \delta^{(1)}(E)\cup(\cup_{k=1}^{s}E^{(1)}_k)
$$
and
\begin{equation}\label{eq_lemma_new_b_2}
\lceil\delta^{(2)}(E)\rceil_{src}\subseteq \lceil\delta^{(1)}(E)\cup(\cup_{k=l}^{s}E^{(1)}_k)\rceil_{src}.
\end{equation}
Let $v_i\rightarrow v_j\in \delta^{(2)}(E)\cap E^{(1)}_l$.
Note that $v_j\in\lceil\delta^{(2)}(E)\rceil_{tar}$. 
According to the definition of Delta-property, we have $v_j$ has at least one $(2,{\bf G})$-follower in $\lceil\delta^{(2)}(E)\rceil_{src}$.
By Eq.  (\ref{eq_lemma_new_b_2}) we have $v_j$ has $(2,{\bf G})$-followers in $\lceil\delta^{(1)}(E)\cup(\cup_{k=l}^{s}E^{(1)}_k)\rceil_{src}$.
However, by $v_j\in\lceil E_{l}^{(1)}\rceil_{tar}$ and the definition of Delta-property, we know that $v_j$ should have no $(2,{\bf G})$-follower in $\lceil\delta^{(1)}(E)\cup(\cup_{t=l}^{s}E^{(1)}_t)\rceil_{src}$, which leads to conflict. Therefore, $\delta^{(2)}(E)\subseteq\delta^{(1)}(E)$. 
Note that $|\delta^{(1)}(E)|\geq |\delta^{(2)}(E)|>0$.
Then, we obtain $\delta^{(1)}(E)\subseteq\delta^{(2)}(E)$ in the same way. 
Consequently, we have $\delta^{(1)}(E)=\delta^{(2)}(E)$.

{Case 2:} $|\delta^{(2)}(E)|=0$. We show $|\delta^{(1)}(E)|=0$ through the reverse proving.
Assume $|\delta^{(1)}(E)|>0$.
According to {Case 1}, we have $\delta^{(1)}(E)\subseteq\delta^{(2)}(E)$.
Thus, $|\delta^{(2)}(E)|\geq |\delta^{(1)}(E)|>0$, which contradicts the assumption that $|\delta^{(2)}(E)|=0$. 
Consequently, we have $|\delta^{(1)}(E)|=0$.
Finally, we obtain $\delta^{(1)}(E)=\delta^{(2)}(E)=\emptyset$.

\section*{Appendix D: Proof of Theorem 1}
Let ${\bf G}={\bf G}(t)$ and $|t|=+\infty$.
We prove Theorem 1 by showing that if all the edges in $E$ share a common label, then $\Delta(E)=1$.
Let $E(0)=E$.
According to Lemma 1, there exists a node $v_{i_1}$ in $\lceil\,E(0)\,\rceil_{tar}$ which has no $(2,{\bf G})$-follower in $\lceil\,E(0)\,\rceil_{src}$.
Let $E(1)=E(0)\setminus \{e\in E(0)|v_{i_1}= \lceil e\rceil_{tar}\}$. 
Obviously, all the edges in $E(1)$ lie in the same layer of ${\bf G}$.
Then by Lemma 1 again, we obtain $v_{i_2}$ in $\lceil\,E(1)\,\rceil_{tar}$ which has no $(2,{\bf G})$-follower in $\lceil\,E(1)\,\rceil_{src}$.
Let $E(2)=E(1)\setminus \{e\in E(1)|v_{i_2}=\lceil e\rceil_{tar}\}$.
Repeat this removal operation on $E$ until all the edges in $E$ are removed.
Finally, we have $\delta(E)=\emptyset$ and $\Delta(E)=1$.

\end{document}